# Design and optimization of CdSe-CuSbSe$_2$-based double-junction two-terminal tandem solar cells with V$_{OC}$> 2.0 V and PCE over 42%

Sheikh Noman Shiddique[1], Ahnaf Tahmid Abir[1], Md Jayed Hossain[2], Mainul Hossain[2], and Jaker Hossain[1*]

[1]Solar Energy Laboratory, Department of Electrical and Electronic Engineering, University of Rajshahi, Rajshahi 6205, Bangladesh.

[2]Department of Electrical and Electronic Engineering, University of Dhaka, Dhaka-1000, Bangladesh.

**Abstract**

In this article, we demonstrate CdSe-CuSbSe$_2$-based double junction two-terminal tandem solar cells simulated with SCAPS-1D. The highest performance of the tandem cell has been confirmed by optimizing the electrical and optical properties of window, top absorber, CdSe (bandgap 1.7 eV), bottom absorber, CuSbSe$_2$ (bandgap 1.08 eV) and back surface layers. In addition, the effect of different parameters such as thickness, doping, defect density of different layers has been investigated in details. With the optimized condition, the modeled CdSe-CuSbSe$_2$ double-junction two-terminal tandem solar cell displays the noticeable efficiency of 42.64% with open circuit voltage of 2.09 V, short circuit current density of 24.09 mA/cm$^2$ and fill factor of 84.36%, respectively. These results are highly propitious for the construction of all-chalcogenide based high performance tandem photovoltaic cells in the future.

# 1. Introduction

Use of solar energy offers a promising solution to battle the world growing energy demands, as well as to overcome the environmental pollution threats produced from the burning of fossil fuel. Single junction photovoltaic (PV) cell with constant bandgap absorbs only a portion of the solar spectrum having energy higher than or equal to the bandgap energy of the absorber layer. Its efficiency is limited to 33% (Shockley Queisser (SQ) limit) by non-absorption and thermalization losses [1]. The first one is attributed to the photons having energy less than the bandgap of the absorber that cannot create electron-hole pairs and the later one is the outcome of the absorption of photons with energy exceeding the bandgap [2]. However, tandem solar cell with multi-junctions (MJs) can overcome this limit as they consist of sub-cells having different bandgaps (the highest one facing the sun) where each cell absorbs various parts of the solar spectrum [3-5]. The theoretical efficiency of MJ tandem solar cells increase with number of junctions e.g. double-junction exhibits ~45% efficiency, triple-junction shows ~51% efficiency and ~55% efficiency for four-junction [6]. Tandem design is accomplished by using monolithically integrated or mechanically stacked, and spectrally split architecture [2].

A double-junction tandem solar cell is the straightforward embodiment of the MJ approach. To get the maximum performance from a double-junction two-terminal tandem PV cell it is necessary to match the photocurrent of the top and bottom cells. This can be achieved by switching the thickness of the absorber layer [7]. The bandgap of the top cell of a thin film tandem cell should be in the range of 1.6 to 1.7 eV and for the bottom cell it should be between 1 and 1.1 eV [8]. One theoretical study shows that with these estimated bandgaps of the top and the bottom cells the tandem cell can achieve an efficiency of 42% in a 2-terminal connection [9]. It is noticeable that the bandgap of the top sub-cell is higher than the bottom cell, because the top cell exploits the higher energy part of the solar spectrum and the longer wavelength photons are absorbed by the bottom cell [5]. The open circuit voltage ($V_{OC}$) is the arithmetic sum of the $V_{OC}$ of each sub-cell [5]. Early studies demonstrate $MAPbI_3$/Si tandems with a conversion efficiency of 13.7%, where the bandgap of $MAPbI_3$ is 1.61 eV [10]. Higher bandgap mixed-halide compositions have also been used to enhance the efficiency [11]. On the other hand,

DSSC/CIGS, CGS/CIGS, Perovskite/CZTS are few experimentally reported works for the fabrication of low cost tandem solar cell [12-14].

In this work, Cadmium selenide (CdSe) and copper antimony selenide (CuSbSe$_2$) chalcogenide semiconductor-based double-junction two terminal tandem cells have been designed and theoretically investigated in details. To get the optimized performance from a double-junction two terminal tandem device, it is necessary to follow the equation, $E_{g\,top}=0.5\times E_{g\,bot} +1.15$ eV, where $E_{g\,top}$ denotes the bandgap of the top absorber layer and $E_{g\,bot}$ presents the bandgap of the bottom absorber layer, to choose the appropriate bandgaps [15]. Therefore, CdSe with a bandgap of 1.7 eV and CuSbSe$_2$ with a bandgap of 1.08 eV have been considered as top and bottom absorber layers, respectively.

CdSe is a semiconducting material with a bandgap of 1.7 eV that belongs to II-VI group. The deposition of CdSe material is achieved through electrodeposition, chemical bath deposition (CBD), and pulsed laser deposition (PLD) [16]. However, due to its small grain size as well as low electrical conductivity, this material needs an annealing and etching treatment so that it can work as a good electrode [17]. As it is generally p-type doped with Li, Na, and P, it is normally used for light-emitting diodes and solar cell applications. Low electrical resistivity and high optical transparency made this material a suitable candidate to employ in solar cell applications [18]. CdSe is cheap, has high carrier mobility, absorbs the entire visible spectrum, high refractive index (~3), low extinction coefficient (~1.0), and a suitable absorption coefficient of $10^4$ cm$^{-1}$.

CuSbSe$_2$ chalcostibite is a p-type narrow bandgap semiconductor that can be grown in a low temperature ranging from 380 to 410 °C and it is more defect tolerant due to having a layered architecture with no dangling bonds [19]. There are different methods to prepare CuSbSe$_2$ material and among them, the most popular processes include electrodeposition, rapid annealing treatment, and hot injection method [20]. Excessive volatile components such as Sb and chalcogen leads to the building up of copper vacancy acceptors, as a result, copper antimony selenide (CuSbSe$_2$) works as a p-type semiconductor [21]. CuSeSb$_2$ is considered to be an ideal absorber because of the presence of 5s$^2$ electrons lone pair [19]. Theoretical studies have shown that CuSbSe$_2$ can reach a power conversion efficiency (PCE) as high as 30% [20]. For the bottom cell, in a two-terminal tandem configuration CuSbSe$_2$ has the appropriate bandgap of 1.08 eV, and a high absorption coefficient of over $10^4$ cm$^{-1}$. Moreover CuSbSe$_2$ can be grown using low-cost methods. Furthermore, the resistivity, carrier concentration, and carrier mobility of $CuSbSe_2$ are 14.4 Ω-cm, 1.33 × $10^{17}$ $cm^{-3}$, and 3.27 $cm^2$/Vs , respectively [21]. Besides, the constituent materials of $CuSbSe_2$ are affluent on earth.

To aggrandize the light-harvesting, the undesirable losses at the junction caused by Fresnel surface reflection must be reduced [22]. Therefore, we need a material like ZnSe as the window layer, which is crystalline for most of the wavelength of the solar spectrum. ZnSe is generally grown by the molecular beam epitaxy method. However, some other methods are also can work as alternatives to produce ZnSe such as thermal evaporation, electro-chemical deposition, metal-organic chemical vapor deposition (MOCVD), and so on [23]. ZnSe material is transparent to high energy photons due to its large energy band gap of 2.7 eV [23-24].

The back surface field (BSF) layer not only plummets the non-uniformity of light absorption but also associates with scattering the photons evenly throughout the absorber layer. We can also accustom the BSF layer as an alternative to the common antireflection coating layer (ARC) to ameliorate the harvesting of light in a solar cell. Another noticeable advantage of a BSF layer is the competency to plunge the surface recombination loss [25-26]. Molybdenum disulphide ($MoS_2$), which can form heterojunctions with both CdSe and $CuSbSe_2$ has been chosen as the back surface field (BSF) layer. Chemical vapor deposition is normally used to grow $MoS_2$. The energy bandgap and diffusion length of this material are 1.6 eV and 1 μm, respectively. Along with this, $MoS_2$ is a good sunlight harvester with an absorption coefficient of $10^4$ $cm^{-1}$. The production cost of this material is also affordable [27-28]. In a previous study, $MoS_2$ was used as the BSF layer with SnS-based solar cells and the recorded theoretical efficiency was about 41% [28].

In this work, numerical computations of a CdSe-$CuSbSe_2$ tandem solar PV cell have been carried out. The material properties like energy bandgap, diffusion length, doping concentration are considered for calculating the device parameters. The change in electrical properties of tandem solar cell configuration with doping concentration, thickness, and bandgap have been investigated in details.

## 2. Device design and simulation methodology

In Figure 1(a), the schematic of the proposed CdSe-CuSbSe$_2$ tandem PV device has been depictured. In the proposed structure, ZnSe and MoS$_2$ have been employs as the window and BSF layers, respectively in the top and bottom cells. In the top cell, CdSe has been employed as the absorber layer and in the bottom cell, the absorber role is played by CuSbSe$_2$. An ideal thin tunnel junction with monolithic architecture has been assumed between the top and bottom cell for the electrical connection. In this assumption, the parasitic absorption and electronic losses have been considered to be negligible or ideal.

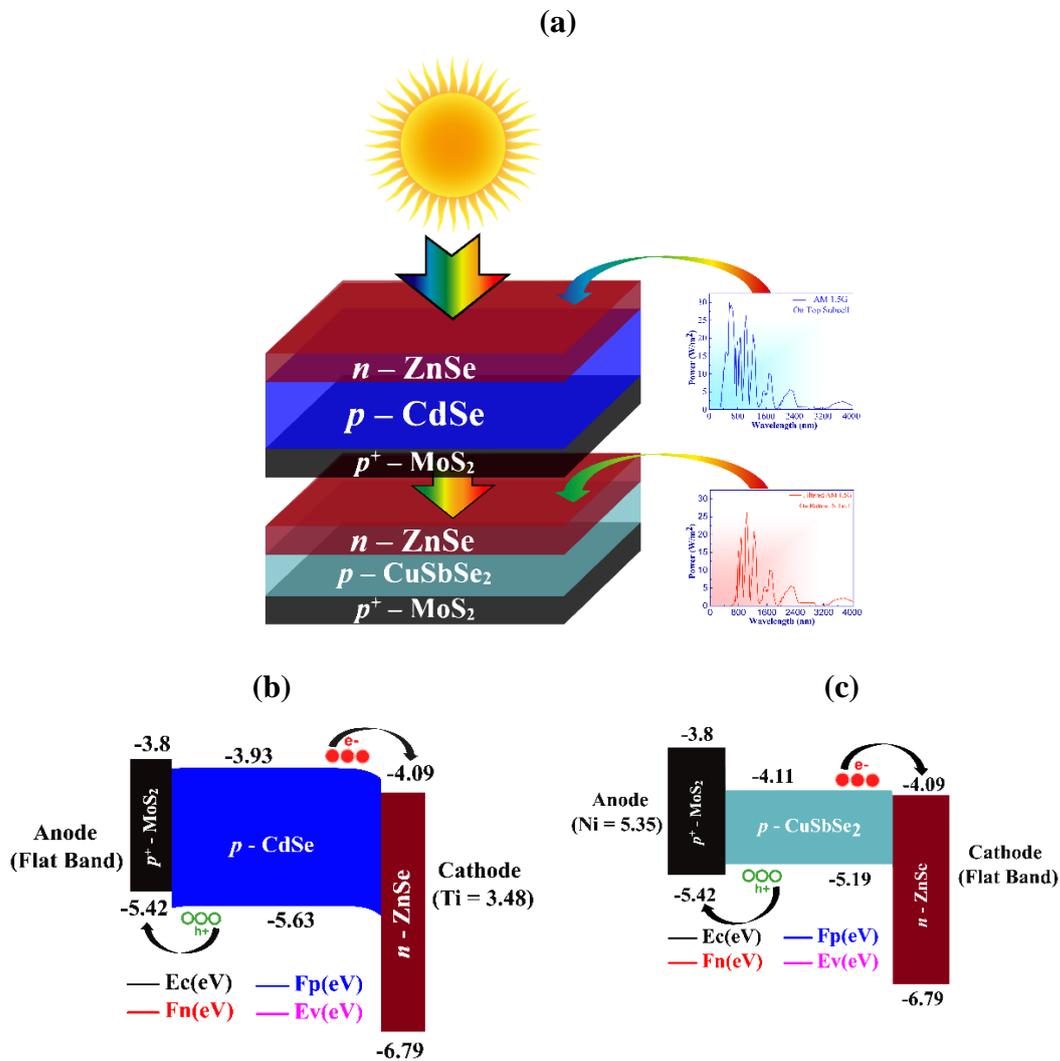

**Figure 1:** (a) Architecture of the tandem cell, (b) energy band diagram of top cell, and (c) energy band diagram of the bottom cell.

Figure 1(b) and (c), show the illumined electronic band structures of the top and bottom cells, respectively. The bandgap and electron affinity of CdSe are 1.7 eV, 3.93 eV, and those of CuSbSe$_2$ are 1.08 eV, 4.11 eV, respectively. Both CdSe and CuSbSe$_2$ have the potential to build a p-n junction with, ZnSe having a bandgap and electron affinity of 2.7 and 4.09 eV, in turn. On the other hand, MoS$_2$ is a material having a bandgap of 1.62 eV and electron affinity of 3.8 eV forms a p-p$^+$ hetero-junction with both CdSe and CuSbSe$_2$ and work as an efficient BSF layer. The lattice parameters of this material may introduce interface defects at the ZnSe/CdSe and ZnSe/CuSbSe$_2$ interfaces due to threading dislocation density (TDD), at the same time, the bulk defects may also be generated during the fabrication period and both of these defects have been considered in the model. The bandgap of CdSe is higher than the bandgap of CuSbSe$_2$ and for this reason, CdSe is chosen as the top material and CuSbSe$_2$ as the bottom cell in tandem architecture.

SCAPS-1D simulation software has been used to compute the performances of the designed CdSe-CuSbSe$_2$ double-junction two terminal tandem solar PV cell with a working temperature of 300 K [29-31]. The simulator solves Poisson equations for electrons and holes for characterizing properties of solar device.

(Poisson's equation) $\quad \frac{\partial^2 \Psi}{\partial x^2} + \frac{q}{\varepsilon}[p(x) - n(x) + N_D - N_A + \rho_p - \rho_n] = 0$ (1)

(Hole continuity equation) $\quad \frac{1}{q}\frac{\partial J_p}{\partial x} = G_{op} - R(x)$ (2)

(Electron continuity equation) $\quad \frac{1}{q}\frac{\partial J_n}{\partial x} = -G_{op} + R(x)$ (3)

where, $\varepsilon$= dielectric constant, q= electron charge, N$_A$= ionized acceptor concentration, N$_D$= ionized donor concentration, J$_p$= hole current, and J$_n$= electron current, respectively. In addition, $\Psi$= electrostatic potential, G$_{op}$ = total carrier generation rate, and R= total recombination rate, p= free hole concentration, and n= free electron density, $\rho_p$= allocation of hole, and $\rho_n$ = allocation of electron, respectively.

The current transport characteristics of holes and electrons in the semiconductor layers can be calculated by the succeeding drift-diffusion equations:

$J_p = -\frac{\mu_p p}{q}\frac{\partial E_{Fp}}{\partial x}$ (4)

$J_n = -\frac{\mu_n n}{q}\frac{\partial E_{Fn}}{\partial x}$ (5)

Standard AM 1.5G sun spectrum has been used to simulate the top cell, on the contrary, in the case of the bottom cell filtered spectrum has been used. This method is vividly utilized to simulate the tandem cells in SCAPS-1D [29-31]. The equation used for filtering the spectrum is,

$$S(\lambda) = S_0 \cdot \exp(\sum_i^n -(a(\lambda) \times d_{mat_i})) \tag{6}$$

Here, the incident AM 1.5 G and the filtered spectrum are expressed by $S_0(\lambda)$ and $S(\lambda)$. An individual material is indicated by $mat_i$ (i=1 for ZnSe, 2 for CdSe and 3 for $MoS_2$). The thickness, absorption coefficient and number of cells are presented by d, $\alpha(\lambda)$ and n respectively.

The following two equations (7) and (8) express the quantum efficiency of the top and the bottom cells in turn [32].

$$QE_{top}(\lambda) = A_{CdSe}(\lambda) \tag{7}$$

$$QE_{bottom}(\lambda) = QE_{CuSbSe_2}(\lambda) \times T(\lambda) \tag{8}$$

Here, A and T indicate the absorption and transmission, respectively. The value of T is found from the ration of $S(\lambda)$ and $S_0(\lambda)$. The quantum efficiency that is taken from the bottom cell separately is represented by $QE_{CuSbSe2}$.

The physical parameters used for ZnSe, CdSe, $CuSbSe_2$ and $MoS_2$ has been taken from literature [16, 19, 28, 33]. The SCAPS optical model sqrt ()-$E_g$ with default settings has been considered for the optical absorption data. Table 1 shows the simulation parameters of CdSe-$CuSbSe_2$-based double-junction tandem solar cell.

**Table 1:** Input parameters used for simulating CdSe-$CuSbSe_2$ Tandem solar cells.

| Parameters | *n*-ZnSe | *p*-CdSe | *p*-CuSbSe$_2$ | *p$^+$*-MoS$_2$ |
|---|---|---|---|---|
| Layer | Window | Absorber | Absorber | BSF |
| Thickness (μm) | 0.2 | 1.00 | 0.871 | 0.2 |
| Bandgap, $E_g$ [eV] | 2.7 | 1.7 | 1.08 | 1.62 |
| Electron affinity, $\chi$ [eV] | 4.09 | 3.93 | 4.11 | 3.8 |

| Parameter | | | | |
|---|---|---|---|---|
| Dielectric permittivity (relative) | 10 | 9.500 | 15.00 | 10.00 |
| Effective DOS at CB [cm$^{-3}$] | $1.5\times10^{18}$ | $2.8\times10^{19}$ | $9.9\times10^{19}$ | $2.8\times10^{19}$ |
| Effective DOS at VB [cm$^{-3}$] | $1.8\times10^{19}$ | $1.2\times10^{19}$ | $9.9\times10^{19}$ | $1.0\times10^{19}$ |
| Electron thermal velocity (cms$^{-1}$) | $1.0\times10^{7}$ | $1.0\times10^{7}$ | $7.3\times10^{6}$ | $1.0\times10^{7}$ |
| Hole thermal velocity (cms$^{-1}$) | $1.0\times10^{7}$ | $1.0\times10^{7}$ | $7.3\times10^{6}$ | $1.0\times10^{7}$ |
| Electron Mobility $\mu_n$ [cm$^2$V$^{-1}$s$^{-1}$] | $5.0\times10^{1}$ | 5.93 | $1.0\times10^{1}$ | $1.2\times10^{1}$ |
| Hole mobility $\mu_P$ [cm$^2$V$^{-1}$s$^{-1}$] | $2.0\times10^{1}$ | $2.5\times10^{1}$ | $1.0\times10^{1}$ | 2.80 |
| Donor density $N_D$ [cm$^{-3}$] | $1.0\times10^{18}$ | 0 | 0 | 0 |
| Acceptor density $N_A$ [cm$^{-3}$] | 0 | $1.0\times10^{16}$ | $1.0\times10^{16}$ | $1.0\times10^{19}$ |
| Type of defect | Single Acceptor | Single Donor | Single Donor | Neutral |
| Energetic distribution | Gaussian | Gaussian | Gaussian | Single |
| Reference for defect energy level $E_t$ | Above $E_V$ (scaps <2.7) | Above $E_V$ (scaps <2.7) | Above $E_V$ (scaps <2.7) | Above $E_V$ (scaps <2.7) |
| Energy level w. r. to Reference (eV) | 1.35 | 0.85 | 0.65 | 0.60 |
| Total defect density, $N_t$ [cm$^{-3}$] | $1.0\times10^{14}$ | $1.0\times10^{14}$ | $1.0\times10^{13}$ | $1.0\times10^{14}$ |
| **Interface input parameters:** | | | | |

| Parameters | Top Cell | | Bottom Cell | |
|---|---|---|---|---|
| | *p*-CdSe/*n*-ZnSe | *p*-CdSe/*p*$^+$-MoS$_2$ | *p*-CuSbSe$_2$/*n*-ZnSe | *p*-CuSbSe$_2$/*p*$^+$-MoS$_2$ |
| Defect type | neutral | neutral | neutral | neutral |
| Total density (1/cm$^2$) | $1.0\times10^{10}$ | $1.0\times10^{10}$ | $1.0\times10^{10}$ | $1.0\times10^{10}$ |

## 3. Results and discussion

The parameters of the sub-cell in the tandem configuration are varied to observe their dominance on the device performance in tandem configuration. These variations will help to get optimized device parameters for maximum efficiency. The device architecture gives higher values of open-circuit voltage V$_{OC}$, short-circuit current J$_{SC}$, and efficiency compared to that of conventional c-Si-based homojunction solar cells [34-37].

### 3.1 Impact of the thickness of both top and bottom cells on PV parameters

Figure 2(a) and (b) delineate the results of changing thickness of *p*-CdSe and *p*-CuSbSe$_2$ absorber layer of the CdSe-CuSbSe$_2$-based double-junction two-terminal tandem device. Figure 2(a) exhibits that with the fluctuation in width of CdSe layer from 0.2 to 1 μm, there is no variation of V$_{OC}$ and it is fixed at 2.09 V. However, further amelioration of thickness may result the enhancement of recombination current that reduces the V$_{OC}$ [38]. On the other hand, J$_{SC}$ is enhanced from 18.63 to 24.08 mA/cm$^2$ in the observed range of thickness. The enhancement of J$_{SC}$ is explicit as more photons are get absorbed with elongation of the absorber layer [39]. The fill factor (FF) changes according to Eq. (9) [40].

$$\text{FF} = \frac{v_{oc} - \ln(v_{oc} - 0.72)}{v_{oc}} \tag{9}$$

with $v_{oc} = \frac{V_{OC}}{nK_BT/q}$,

where, n states the diode ideality factor, $K_B$ represents the Boltzmann constant, T stands for the absolute temperature, q denotes the electronic charge.

It is noticed from the figure that fill factor decreases from 90 to 84%. This can be explained with the increase in diode ideality factor, n which indicates the prevalent recombination mechanisms in the device. The increase in n reveals that the recombination in the depletion region become dominant with the increase in CdSe thickness [40-41]. However, the PCE of the tandem device exhibits an enhancement from 35 to 42.6% due to the increase in $J_{SC}$. An optimized thickness of 1 μm has been considered for the CdSe layer for further probes resulting in the tandem cell PCE of 42.64% with $V_{OC}$ of 2.09 V and $J_{SC}$ of 24.09 mA/cm$^2$, respectively.

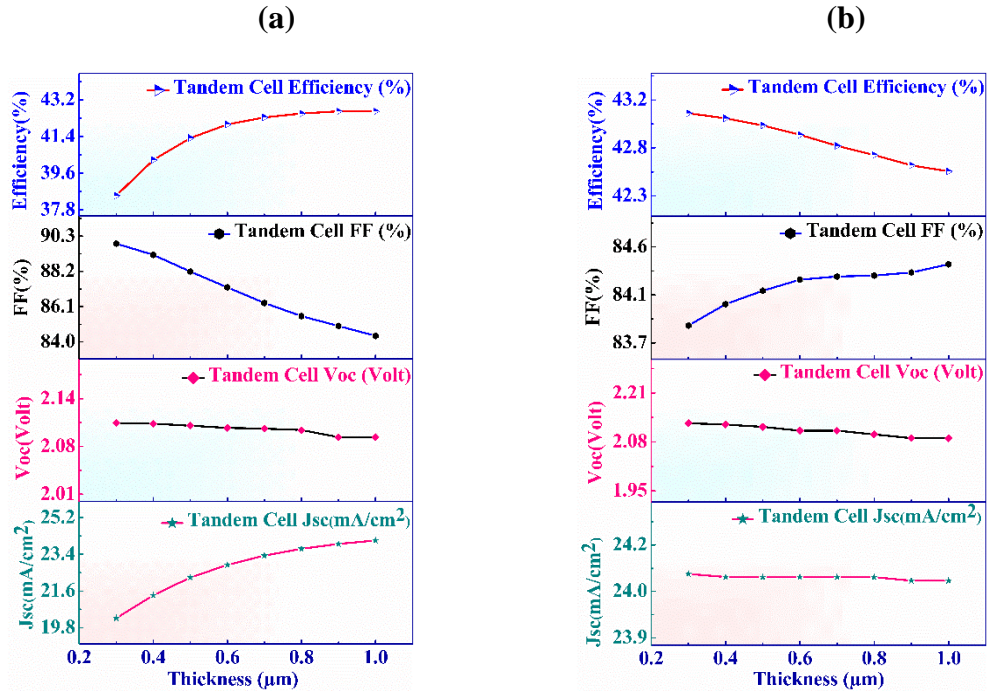

**Figure 2:** (a) Thickness vs. output parameters of top cell and (b) thickness vs. output parameters of bottom cell of the CdSe-CuSbSe$_2$ tandem solar cell.

Figure 2(b) illustrates the impression of CuSbSe$_2$ absorber layer thickness on the PV performance varied from 0.2 to 1 µm. It is observed from the figure that with the advance in absorber thickness, the V$_{OC}$ downfalls inappreciably. The V$_{OC}$ exhibits a value of approximately 2.09 V. However, a constant J$_{SC}$ of 24.09 mA/cm$^2$ is achieved. In the case of the fill factor, we have experienced almost one percent increase from about 83.83 to 84.43% indicating the decrease in diode ideality factor and carrier recombination moves to neutral region [40]. The PCE is slightly reduced from around 43 to 42.5 % due to the fall in open circuit voltage. It can be summarize from these outcomes that as the width of the bottom absorber layer elongates, the output of the PV cell reduces meagerly. The thickness of the bottom absorber layer has been maintained at 871 nm for further studies which has been found from current matching between the top and bottom sub-cell of the CdSe-CuSbSe$_2$ double-junction two terminal tandem solar cell as shown in Table 1.

**3.2 Impact of the doping concentration of both top and bottom cells on PV parameters**

The doping density in CdSe and CuSbSe$_2$ layers has been altered in the range from $1\times 10^{14}$ to $1\times 10^{18}$ cm$^{-3}$ for both the top cell and bottom cell of the CdSe-CuSbSe2 tandem solar cell which are shown in Figure 3(a) and (b), respectively.

Figure 3(a) shows the variation of output performance of the tandem cell with change in doping in top CdSe absorber layer. It is noticed that the J$_{SC}$ decreases rapidly from 24.74 to 11.29 mA/cm$^2$ with increase in doping concentration. This is due to the recombination losses which are higher at higher doping concentrations, leading to the rapid drop in J$_{SC}$ [42-43]. The value of V$_{OC}$ shows an upward trend and increases from 2.09 to 2.13 V. Up to $10^{17}$ cm$^{-3}$, the value of FF decreases from 89 to 79, and reaches 82% at $1 \times 10^{18}$ cm$^{-3}$ indicating that recombination is dominant in the neutral region of the diode at higher doping [41]. Nevertheless, efficiency of the tandem cell suddenly falls from 46.02 to approximately 19.79% when the doping levels increased from $1 \times 10^{14}$ to $1 \times 10^{18}$ cm$^{-3}$ owing to the decrease in $J_{SC}$.

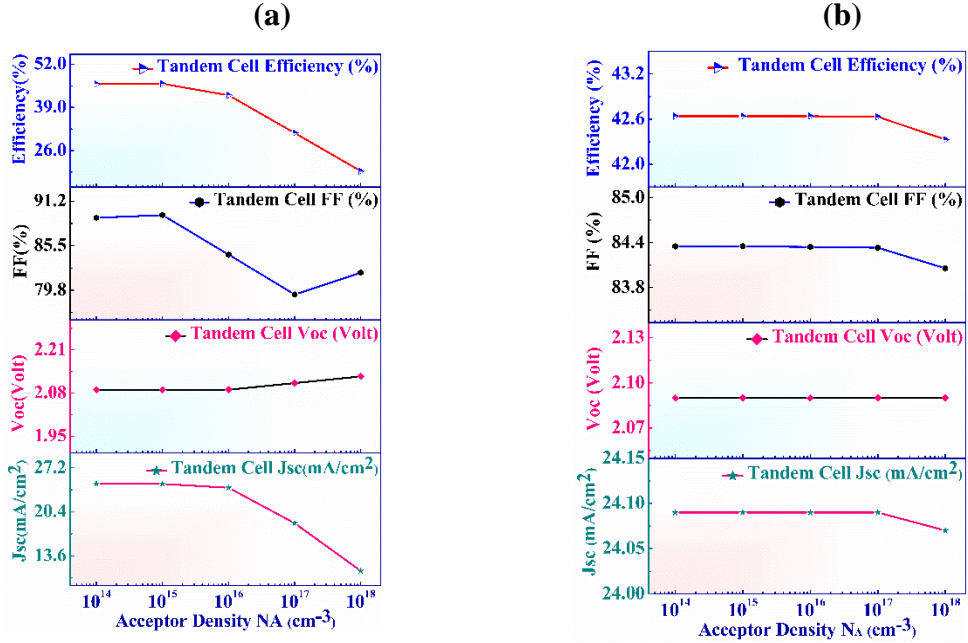

**Figure 3**: (a) Acceptor density of top cell vs. output parameters and (b) acceptor density of bottom Cell vs. output parameters of CdSe-CuSbSe$_2$ tandem solar cell.

Figure 3(b) depicts the output PV performance of CdSe-CuSbSe$_2$ tandem cell with the variation in doping of CuSbSe$_2$ bottom absorber layer. It is revealed from the figure that both the $V_{OC}$ and $J_{SC}$ values are constant at the value of about 2.09 V and 24.09 mA/cm$^2$, respectively when the doping density of the bottom cell absorber layer varies from the optimum value. FF and hence PCE of the device slightly decrease, indicating that the device performance will show a degradation if the dopant is increases in the bottom absorber layer.

### 3.3 Impact of the defect density of both top and bottom cells on PV parameters

The output performance of the solar PV cell largely depends on defects of the different cell layers [44]. In this modeled tandem cell, the total defects in CdSe of the top cell has been altered in the range of $10^{12}$–$10^{18}$ cm$^{-3}$, whereas in CuSbSe$_2$ layer of the bottom cell it has been varied in the range of $10^{11}$–$10^{15}$ cm$^{-3}$.

The undulation of PV performance of CdSe-CuSbSe$_2$ tandem solar cell with total defects of the CdSe top cell is visualized in Figure 4(a). There is a decrease in the value of $V_{OC}$ from around 2.09 to 1.699 V which is happened owing to the rise in the reverse saturation current with higher defects [45- 46]. The $J_{SC}$ has exhibited almost a constant value up to $10^{16}$ cm$^{-3}$ with a slight change from 24.09 to 24.46 mA/cm$^2$ with the increased defects density. Nevertheless, after $10^{16}$ cm$^{-3}$ of defect density, the short circuit current density, $J_{SC}$ has changes from 24.50 mA/cm$^2$ to almost 0 mA/cm$^2$ at a defect of $10^{18}$ cm$^{-3}$.

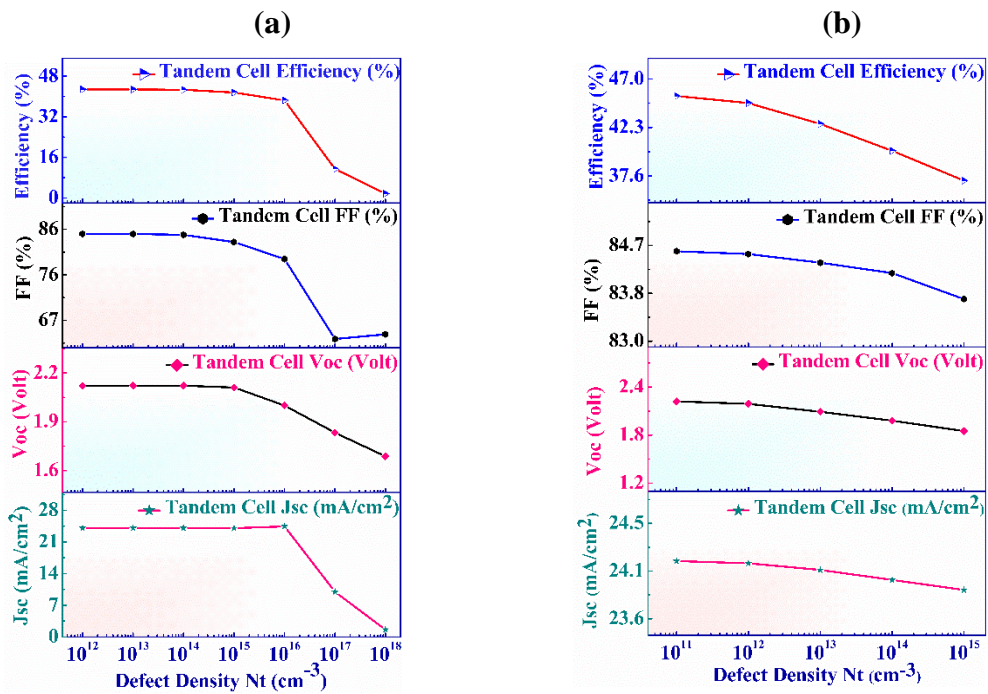

**Figure 4:** (a) Defect density of top cell vs. output parameters and (b) defect density of bottom cell vs. output parameters of CdSe-CuSbSe$_2$ tandem solar cell.

A dip in FF from about 85 to 60% has been observed between the defect orders of $10^{16}$–$10^{18}$ cm$^{-3}$. After crossing the defect of $10^{18}$ cm$^{-3}$, the FF get to be levelled off. The PCE follows almost the path of FF as it decreases from around 42 to 15% in the range between $10^{16}$ and $10^{18}$ cm$^{-3}$.

Figure 4 (b) shows the impression of defects in CuSbSe$_2$ layer on the output of CdSe-CuSbSe$_2$ tandem device. It is noticed from the figure that the $V_{OC}$ degrades from 2.22 to 1.85 V when the defects in CuSbSe$_2$ bottom layer changes from $1 \times 10^{11}$ to $1 \times 10^{15}$ cm$^{-3}$. It is concluded that as the defect density is enhanced it increases the recombination of photocarriers in the CuSbSe$_2$ absorber layer and at the same time decrease the carrier lifetime which deteriorates the output performance of the tandem device. The rise in recombination centers in CuSbSe$_2$ layer also promote the reduction in shunt resistance that also minimize the $V_{OC}$ of the solar PV cell [47]. However, the decrease in $J_{SC}$ from 24.17 to 23.91 mA/cm$^2$ is significant. Consequently, the FF follows a zig-zag path and falls from 84.56 to 83.73% and the PCE declines from 45.35 to 37.14%.

### 3.5 Window and BSF layer effect on tandem solar cell

### 3.5.1 Window layer impact on tandem cell

Figure 5 delineates the role of the variation of various factors in the ZnSe window layer of top cell.

Figure 5(a) describes the impact of variation of thickness of ZnSe window layer on the performance of tandem cell. The thickness has been altered in the range of 0.2 to 1 μm. It is noticed from the figure that the $V_{OC}$ of cell is almost constant at 2.09 V whereas $J_{SC}$ slightly varies from 24.08 to 24.03 mA/cm$^2$, FF varies from 84.36 to 84.73% and hence efficiency varies from 42.64 to 42.74%.

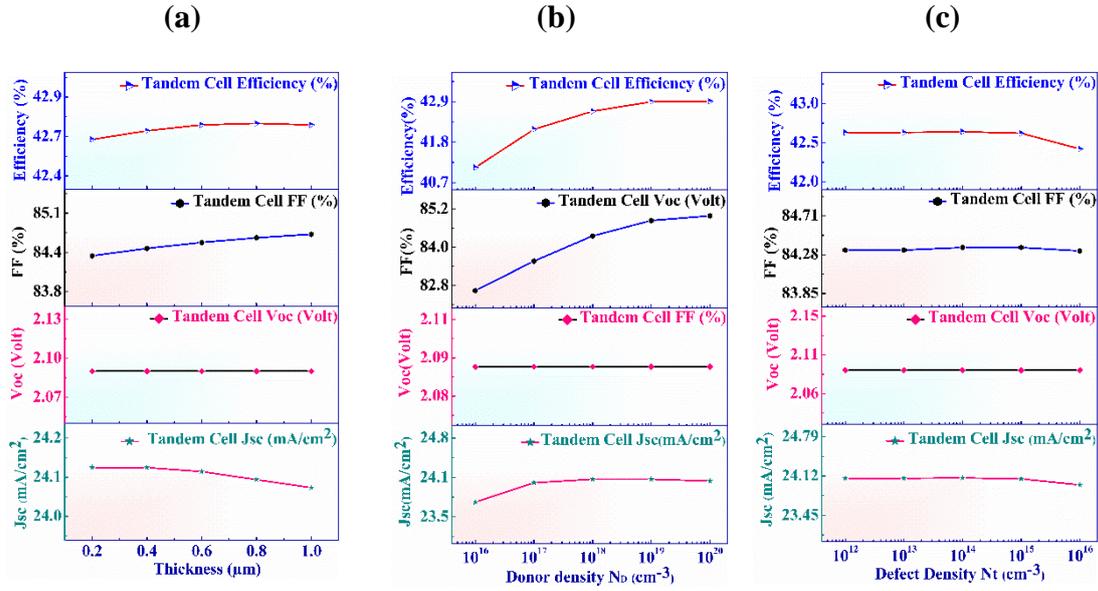

**Figure 5:** Efficiency, FF, $V_{OC}$, $J_{SC}$ variation of CdSe-CuSbSe$_2$ tandem solar cell with top cell's window (a) layer thickness, (b) doping, $N_D$, and (c) defect, $N_t$.

The effect of the variation of acceptor density ($N_D$) of window layer (ZnSe) of top cell in the outcomes of tandem cell are shown in Figure 5(b). The doping has been varied from $10^{16}$ to $10^{20}$ cm$^{-3}$. It can be noted that the Voc of the tandem cell is constant at 2.09 V, Jsc increases from 23.7 to 24.06 mA/cm$^2$, FF rises from 82.64 to 84.99% and therefore efficiency increases from 41.11 to 42.91%, respectively.

Figure 5(c), depictures the tandem cell output by varying defect density ($N_t$) of top cell window layer. The defects have been varied from $10^{12}$ to $10^{16}$ cm$^{-3}$. It is noticed from the figure that all the output parameters are almost constant with the variation of defects in the ZnSe layer.

Figure 6(a) reveals the fluctuations of tandem cell output with varying thickness of bottom cell's ZnSe window layer from 0.2 to 1 μm. As seen in the figure, $V_{OC}$ is constant at 2.09 volt, $J_{SC}$ is constant at 24.09 mA/cm$^2$, FF falls from 84.36 to 84.26% and therefore, efficiency slightly downs from 42.64 to 42.51%, respectively.

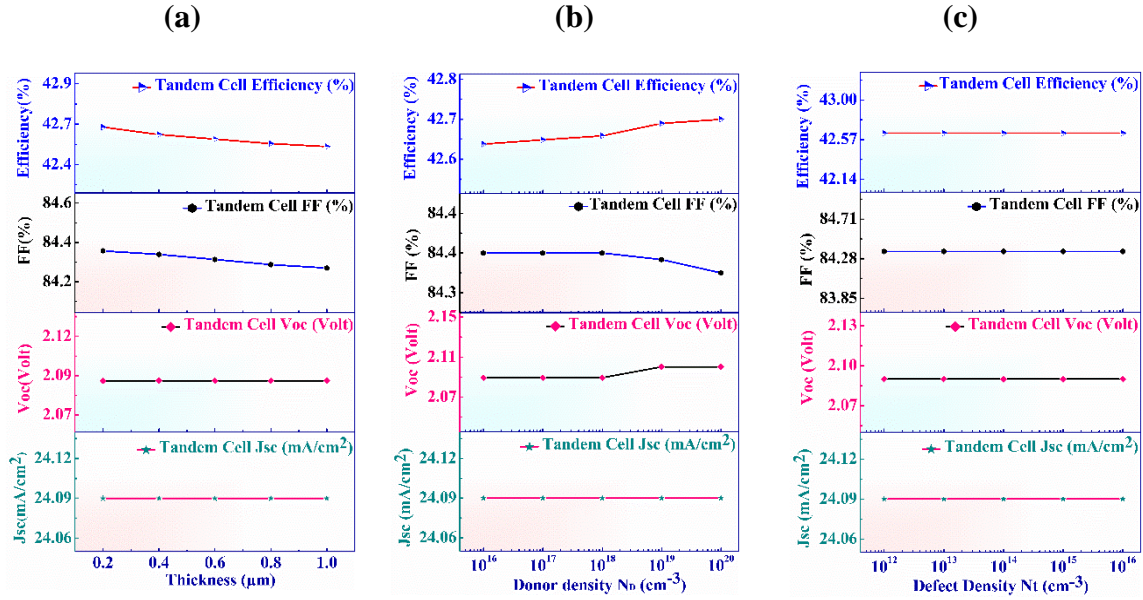

**Figure 6:** Efficiency, FF, $V_{OC}$, and $J_{SC}$ variation of CdSe-CuSbSe$_2$ tandem solar cell with bottom cell's window (a) layer thickness, (b) acceptors, $N_D$, and (c) bulk defects, $N_t$.

Figure 6(b), exhibits the cell output variation with acceptor density, $N_D$ of top cell ZnSe window layer from $10^{16}$ to $10^{20}$ cm$^{-3}$. We see that tandem cell yields a constant $V_{OC}$ of 2.09 volt, $J_{SC}$ of 24.09mA/cm$^2$. The FF slightly increase to a marginal value from 84.36 to 84.33% and efficiency slightly varies from 42.62 to 42.68%, respectively.

Figure 6(c) delimitates the impact of varying defect density, $N_t$ of top cell ZnSe window from $10^{12}$ to $10^{16}$ cm$^{-3}$, and it is seen from the figure that almost all of the output parameters are constant within this range of defects.

The carrier mobility and diffusion length is higher for the ZnSe semiconductor and therefore, the device performance dose not alters much with the alteration of thickness, doping and defects of ZnSe layers [48].

### 3.5.2 MoS$_2$ BSF layer impact on tandem cell

Herein, the effects of the change in the parameters of MoS2 BSF layer have been describes in depth. Figure 7(a) depictures the role of varying thickness of top cell MoS$_2$ BSF from 0.2 to 1 μm. The computation provides the Voc = 2.09 volt, $J_{SC}$ in the range from 24.08 to 24.76

mA/cm2, FF from 84.36 to 84.22% and therefore efficiency varies in the range from 42.64 to 43.78%, respectively for the tandem cell.

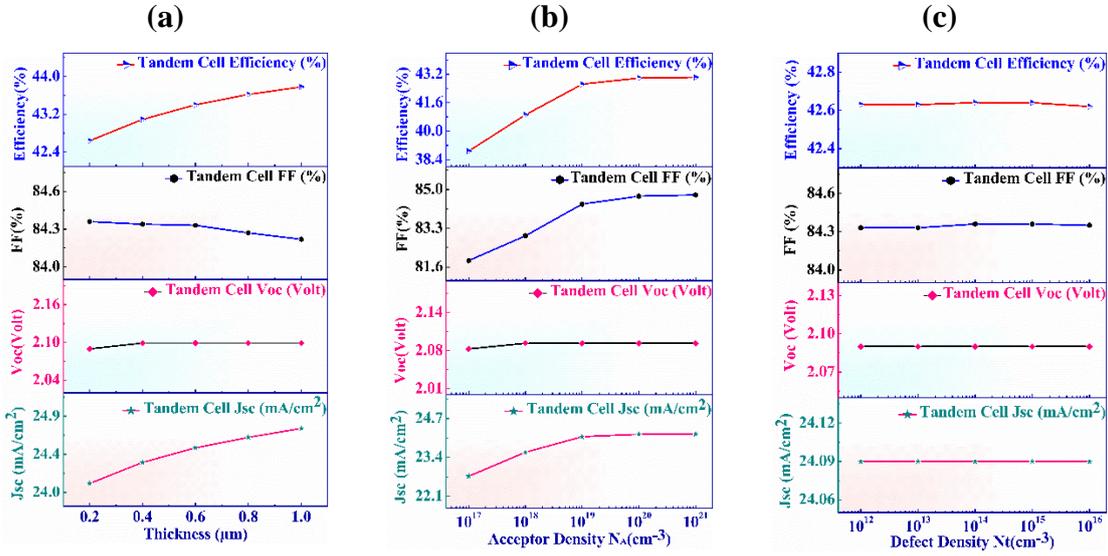

**Figure 7:** Efficiency, FF, $V_{oc}$, $J_{sc}$ variation of CdSe/CuSbSe$_2$ tandem solar cell with of top cell's BSF (a) layer thickness, (b) doping, $N_A$ (c) defects, $N_t$.

It is visualized in Figure 7(b) that Voc changes from 2.08 to 2.09 volt, Jsc enhances from 22.77 to 24.18 mA/cm2 , FF varies from 81.88 to 84.76% and efficiency rises from 38.88 to 43.03% respectively owing to the change in acceptor density ($N_A$) of MoS$_2$ BSF layer of top cell from $10^{17}$ to $10^{21}$ cm$^{-3}$.

The role of varying defect density ($N_t$) of MoS$_2$ BSF layer of top cell from $10^{12}$ to $10^{16}$ cm$^{-3}$ is shown in Figure 7(c) that depicts that it has nearly no impact on the device performance.

The output performance of the tandem cell slightly varies with the variation in MoS$_2$ layer in the top cell especially the current with doping that may be result due to the decrease in surface recombination velocity with doping that enhances built-in voltage at the junction [49-50].

The performance of CdSe-CuSbSe$_2$ tandem solar cell with respect to the change in thickness, doping and defects in MoS$_2$ layer in bottom cell are delineates in Figure 8(a)-(c). It can be noticed from the figure that the tandem cell performance does not significantly varies with the variations in these parameters in back surface layer in the bottom cell.

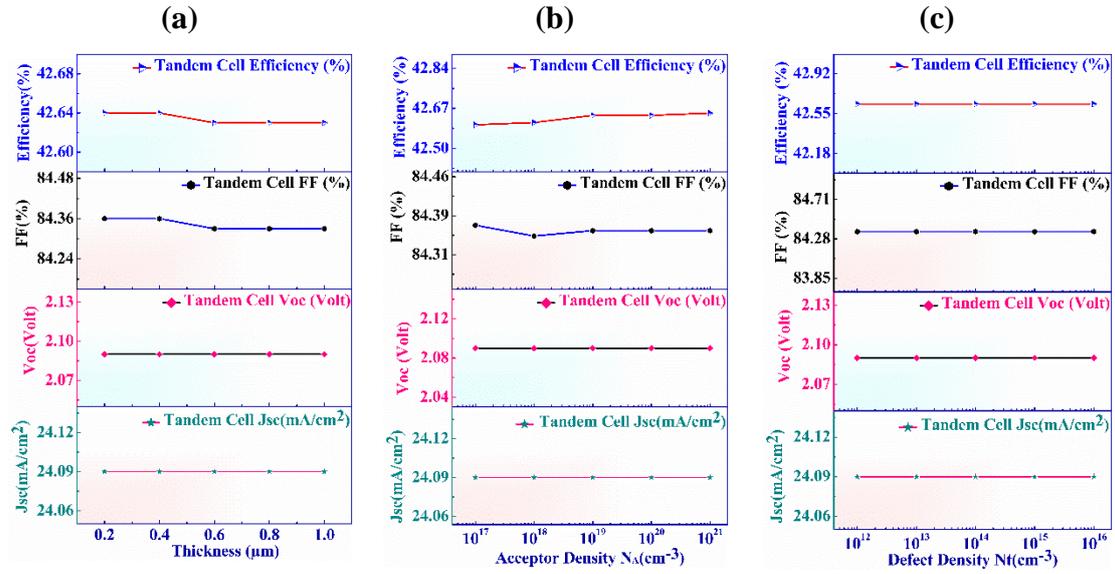

**Figure 8:** Efficiency, FF, $V_{OC}$, and $J_{SC}$ variation of CdSe/CuSbSe$_2$ tandem solar cell with bottom cell's BSF (a) layer thickness, (b) doping, Na and (c) defects, $N_t$.

### 3.6 Optimized device performance

Figure 9(a) depicts the J-V curves of the top, the bottom and the tandem cells CdSe-CuSbSe$_2$ tandem solar cell. It is most important to match the currents in top and bottom cell for attaining the maximum performance of two terminal tandem solar cells. Therefore, current matching technique has been performed wherein, the filtered spectrum by the CdSe top cell has been utilized to illuminate the CuSbSe$_2$ bottom cell with varying thickness of the bottom cell to make current equal to that of top cell. The bottom cell has the lower bandgap, current value will be higher due to availability of large spectrum thus bottom absorber thickness needs to be reduced. At current matching point, optimum value of the width for bottom CuSbSe$_2$ absorber layer is 0.871 μm and that for top CdSe absorber layer is 1 μm and the matched short circuit current is 24.09 mA/cm$^2$. In optimized condition, the thickness for both the window and BSF layer are 0.2 μm. The optimized defect density of the window, the top absorber, the bottom absorber and the BSF layer are $10^{14}$, $10^{14}$, $10^{13}$ and $10^{14}$ cm$^{-3}$ respectively. At the same time, the optimized doping concentrations are $10^{18}$, $10^{16}$, $10^{16}$, and $10^{19}$ cm$^{-3}$ for the window, the top absorber, the bottom absorber and the BSF layer, respectively.

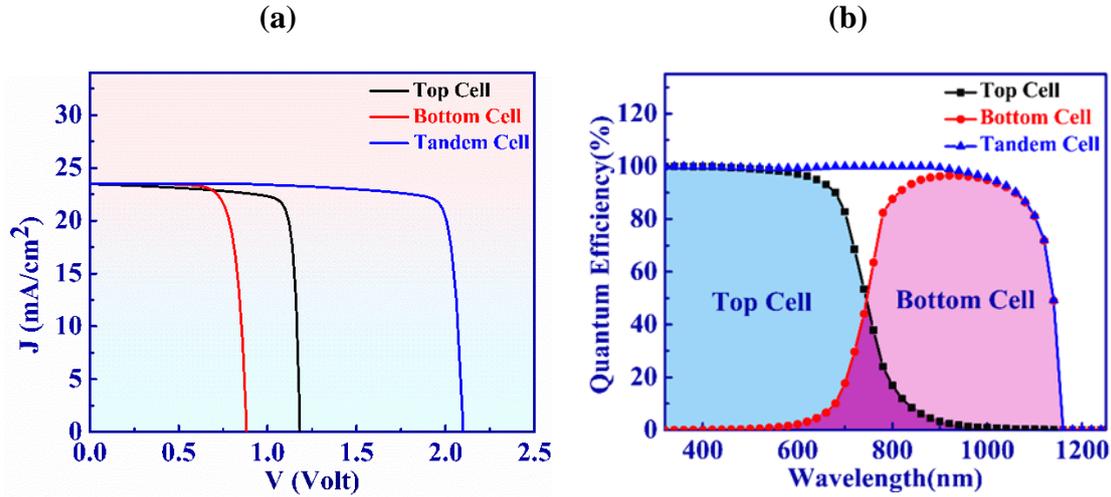

**Figure 9:** (a) J-V and (b) quantum efficiency (QE) curves of CdSe-CuSbSe$_2$ tandem solar PV cell.

Figure 9(b) delineates the fluctuation in quantum efficiency, QE with Wavelength of the incident solar spectrum of the CdSe-CuSbSe$_2$ tandem solar cell. The QE measures the ratio of photogenerated carriers to total photons impinged on the cell. This expresses that the QE will be 100% in the case of every photons absorbed by the PV cell. In general, the QE is computed with the change in wavelength (λ) or energy of the incident light. However, QE is 100% only for a particular wavelength of the light [50]. It is observed from the figure that QE is almost 90% at a wavelength of 700 nm for the CdSe top absorber layer and it suddenly falls to about 10% at 800 nm. On the other hand, when the filtered spectrum falls on the CuSbSe$_2$ sub cell, the QE is about 15% at the wavelength of 700 nm. The QE becomes highest to about 98% at the wavelength of about 1000 nm, then it becomes zero at the wavelength of ~1150 nm. Therefore, it can be inferred that CdSe-CuSbSe2 tandem solar cell exhibits an excellent QE in the visible and infra-red region up to the wavelength of 1150 nm.

## 4. Conclusion

A detailed theoretical investigation on a highly efficient CdSe-CuSbSe$_2$ chalcogenide-based double-junction two-terminal tandem solar cell has been accomplished. Thickness of the absorber layers in both the top and the bottom cells is optimized to obtain current matching condition. The role of doping, defects in different layers such as CdSe top and CuSbSe$_2$ bottom absorber layers, ZnSe window and MoS$_2$ BSF layers has been probed in details. The simulated results show large value of open circuit voltage of 2.09 V and high efficiency of 42.64% can be achieved with J$_{SC}$ of 24.09 mA/cm$^2$ and FF of 84.36%. Experimental realization of such work could possibly lead to highly efficient and low cost two-terminal tandem solar cells.

## Data Availability

The data used to support the findings of this study are available from the corresponding author upon request.

## Consent

This article has the consent from all the authors.

## Authors' Contributions

**S. N. Shiddique** (Data curation: Equal; Formal analysis: Equal; Investigation: Equal; Writing – original draft: Equal). **A. T. Abir** (Data curation: Equal; Formal analysis: Equal; Investigation: Equal; Validation: Equal; Writing – original draft: Equal). **M. J. Hossain** (Data curation: Equal; Formal analysis: Equal; Investigation: Supporting; Writing – original draft: Equal). **M. Hossain** (Formal analysis: Supporting; Methodology: Supporting; Supervision: Supporting; Writing – review & editing: Equal). **J. Hossain** (Conceptualization: Lead; Data curation: Equal; Formal analysis: Equal; Supervision: Lead; Validation: Lead; Writing – original draft: Equal; Writing – review & editing: Lead).

**Conflicts of Interest**

The authors have no conflicts of interest.

**Corresponding Authors:**

*E-mail: jak_apee@ru.ac.bd (Jaker Hossain).

**Acknowledgements**

This study was supported by the University of Rajshahi and University of Dhaka, Bangladesh. The authors are indebted to Prof. Dr. Marc Burgelman, University of Gent, Belgium, for supporting with the SCAPS simulation software.